\documentclass[11pt,english]{article}
\usepackage[a4paper]{geometry}
\usepackage{amsmath,latexsym,amsxtra,enumerate,color,cancel,amsthm}
\usepackage{enumitem}
\usepackage{amsthm}
\usepackage[utf8]{inputenc}
\usepackage[english]{babel}
\usepackage[usenames,dvipsnames,svgnames,table]{xcolor}
\usepackage{amsfonts, amssymb,  amscd, graphicx,stmaryrd}
\usepackage{mathrsfs}
\usepackage{algorithm, algorithmicx, algpseudocode}
\usepackage{graphicx}
\usepackage{bbm,bbding}
\usepackage[active]{srcltx}
\usepackage{xspace}

\usepackage{aliascnt}
\RequirePackage{hyperref}
   
\newcommand{\imagewidth}{8cm}

\newcommand{\E}{\mathbb{E}}

\newcommand{\R}{\mathbb{R}}
\newcommand{\Rd}{\mathbb{R}^d}

\newcommand{\cF}{\mathcal{F}}

\newcommand{\cL}{\mathcal{L}}
\newcommand{\equa}{\begin{eqnarray*}}
\newcommand{\tion}{\end{eqnarray*}}
\newcommand{\equal}{\begin{eqnarray}}
\newcommand{\tionl}{\end{eqnarray}}
\newcommand{\ind}{\mathbbm{1}}

\newcommand{\f}{\mathrm{f}}

\newcommand{\trend}{a}

\newcommand{\processtn}[1]{#1_{t_n}}
\newcommand{\processtnplus}[1]{#1_{t_{n+1}}}
\newcommand{\Xtn}{\processtn{X}}
\newcommand{\Xtnplus}{\processtnplus{X}}

\newcommand{\Loperator}[1]{\frac{1}{2}\sum_{i,j=1}^d(\sigma \sigma^\top)_{i,j}(x) \frac{\partial^2}{\partial x_i \partial x_j} + \sum_{i=1}^d \mu_i(x) \frac{\partial}{\partial x_i}}

\newcommand{\laplace}{\Delta}
\newcommand{\grad}{\nabla}

\theoremstyle{plain}

\newcommand{\registerTheorem}[2]{
	\newaliascnt{#1}{theorem}
	\newtheorem{#1}[#1]{#2}
	\expandafter\providecommand\expandafter
	*\csname#1autorefname\endcsname{#2}
	\aliascntresetthe{#1}
}

\newtheorem{theorem}{Theorem}[section]

\registerTheorem{thm}{Theorem}
\registerTheorem{lem}{Lemma}
\registerTheorem{lemma}{Lemma}
\registerTheorem{prop}{Proposition}
\theoremstyle{definition}
\registerTheorem{rem}{Remark}
\registerTheorem{defi}{Definition}
\registerTheorem{cor}{Corollary}
\registerTheorem{ass}{Assumption}

\title{
	A deep neural network algorithm for semilinear elliptic PDEs\\
	with applications in insurance mathematics
}

\author{ Stefan Kremsner$^1$   \hspace{1.0em}   Alexander Steinicke$^2$   \hspace{1.0em}   Michaela Szölgyenyi$^3$}

\date{Preprint, December 2020}
\begin{document}
\parindent 0pt

\maketitle


\begin{abstract}
In insurance mathematics, optimal control problems over an infinite time horizon arise when computing risk measures. An example of such a risk measure is the expected discounted future dividend payments. In models which take multiple economic factors into account, this problem is high-dimensional.
The solutions to such control problems correspond to solutions of deterministic semilinear (degenerate) elliptic partial differential equations. In the present paper we propose a novel deep neural network algorithm for solving such partial differential equations in high dimensions in order to be able to compute the proposed risk measure in a complex high-dimensional economic environment.
The method is based on the correspondence of elliptic partial differential equations to backward stochastic differential equations with unbounded random terminal time. In particular, backward stochastic differential equations which can be identified with solutions of elliptic partial differential equations are approximated by means of deep neural networks.
\end{abstract}


{\noindent \textit{Keywords:} Backward stochastic differential equations, semilinear elliptic partial differential equations, stochastic optimal control, unbounded random terminal time, machine learning, deep neural networks.\\
\noindent \textit{Mathematics Subject Classification (MSC 2020):} 60H35, 65N75, 68T07
}
{\noindent
\footnotetext[1]{Department of Mathematics, University of Graz, Heinrichstra\ss{}e 36, 8010 Graz, Austria. \hspace*{1.5em}
 {\tt stefan.kremsner{\rm@}uni-graz.at} \Envelope}
\footnotetext[2]{ Department of Mathematics and Information Technology, Montanuniversitaet Leoben, Peter Tunner-Stra\ss e 25/I, 8700 Leoben, Austria. \hspace*{1.5em}  {\tt  alexander.steinicke{\rm@}unileoben.ac.at}}}
\footnotetext[3]{Department of Statistics, University of Klagenfurt, Universit\"atsstra\ss{}e 65-67, 9020 Klagenfurt, Austria. \hspace*{1.5em}  {\tt  michaela.szoelgyenyi{\rm@}aau.at}}


\section{Introduction}

Classical optimal control problems in insurance mathematics include finding risk measures like the probability of ruin or the expected discounted future dividend payments.
Mathematically, these are problems over a potentially infinite time horizon, ending at an unbounded random terminal time -- the time of ruin of the insurance company.
In recent models which take multiple economic factors into account, the problems are high dimensional.
For computing these risk measures, optimal control problems need to be solved numerically.
A standard method for solving control problems is to derive the associated Hamilton-Jacobi-Bellman (HJB) equation -- a semilinear (sometimes integro) partial differential equation (PDE) and show that its (numerical) solution also solves the original control problem.
In the case of infinite time horizon problems, these HJB equations are (degenerate) elliptic.
\emph{In this paper we propose a novel deep neural network algorithm for semilinear (degenerate) elliptic PDEs associated to infinite time horizon control problems in high dimensions.}\\

We apply this method to solve the dividend maximization problem in insurance mathematics.
This problem originates in the seminal work by De Finetti \cite{definetti1957}, who introduced expected discounted future dividend payments as a valuation principle for a homogeneous insurance portfolio.
This constitutes an alternative risk measure to the (even more) classical probability of ruin.
Classical results on the dividend maximization problem are \cite{shreve1984,jeanblanc1995,radner1996,asmussen1997}.
Overviews can be found in \cite{albrecher20092,avanzi2009}, for an introduction to optimization problems in insurance we refer to \cite{schmidli2008,azcue2014}.
Recent models for the surplus of an insurance company allow for changes in the underlying economy. Such models have been studied, e.g., in \cite{jiang2012,sotomayor2011,zhu2013,LST14,S13,S16,reppen20}.
In \cite{LST14, S13, S16} hidden Markov models for the underlying economic environment were proposed that allow for taking (multiple) exogenous, even not directly observable, economic factors into account. While these authors study the dividend maximization problem from a theoretical perspective, we are interested in \emph{computing} the risk measure. However, classical numerical methods fail when the problem becomes high-dimensional, that is for example when exogenous economic factors are taken into account.
In this paper we propose a novel deep neural network algorithm to solve high-dimensional problems. As an application we use it to solve the dividend maximization problem in the model from \cite{S16} in high dimensions numerically.\\

Classical algorithms for solving semilinear (degenerate) elliptic PDEs like finite difference or finite element methods suffer from the so-called curse of dimensionality -- the computational complexity for solving the discretized equation grows exponentially in the dimension. In high-dimensions (say $>10$) one has to resort to costly quadrature methods such as multilevel-Monte Carlo or the quasi-Monte Carlo-based method presented in \cite{KLST19}. In recent years, deep neural network (DNN) algorithms for high-dimensional PDEs have been studied extensively. Prominent examples are \cite{han2018solving,weinan2017deep}, where semilinear parabolic PDEs are associated with backward stochastic differential equations (BSDEs) through the (non-linear) Feynman-Kac formula and a DNN algorithm is proposed that solves these PDEs by solving the associated BSDEs.
In the literature there exists a variety of DNN approaches for solving PDEs, in particular (degenerate) parabolic ones. Great literature overviews are given, e.g., in \cite{grohs2019deep, beck2020overcomingelliptic}, out of which we list some contributions here:
\cite{
	beck2019deepsplitting,
	Kolmogorov,
	BeckJentzenE2019,
	BeckerCheridito2019,
	BeckerCheriditoJentzen2019,
	Berg2018AUD,
	chan2019machine,
	chen2019deep,
	Dockhorn2019,
	weinan2018deep,
	Farahmand2017DeepRL,
	FujiiTakahashi2019,
	GoudenegeMolent2019,
	han2018convergence,
	han2020derivativefree, 
	henry2017deep,
	HurePhamWarin2019,
	JacquierOumgari2019,
	LongLuMaDong2018,
	lu2019deepxde,
	LyeMishraRay2019,
	magill2018neural,
	PhamWarin2019,
	Raissi2018DeepHP,
	SirignanoDGM2017}.\\
	
While in mathematical finance control problems (e.g., investment problems) are studied over relatively short time horizons, leading to (degenerate) parabolic PDEs, in insurance mathematics they are often considered over the whole lifetime of the insurance company, leading to (degenerate) elliptic PDEs.
For elliptic PDEs, a multi-level Picard iteration algorithm is studied in \cite{beck2020overcomingelliptic}, a derivative-free method using Brownian walkers without explicit calculation of the derivatives of the neural network is studied in \cite{han2020derivativefree}, and a walk-on-the-sphere algorithm is introduced in  \cite{grohs2020deepneuralelliptic} for the Poisson equation, where the existence of DNNs that are able to approximate the solution to certain elliptic PDEs is shown.\\

In the present article we propose a novel DNN algorithm for a large class of semilinear (degenerate) elliptic PDEs.
For this, we adopt the approach from \cite{han2018solving} for (degenerate) parabolic PDEs. The difference here is that we use the correspondence between the PDEs we seek to solve and BSDEs with random terminal time. This correspondence was first presented in \cite{pardoux1998backward}, and elaborated afterwards, e.g., in \cite{darling1997backward,briand1998backward,pardoux1999backward,royer2004backward,confortola2007backward}.

As these results are not as standard as the BSDE correspondence to parabolic PDEs, we summarize the theory in Section \ref{sec:bsde-pde} for the convenience of the reader. In Section \ref{sec:algorithm} we present the DNN algorithm, and test it in Sections \ref{sec:Poisson} and \ref{sec:quadratic}.
In Section \ref{sec:dividends} we present the model from \cite{S16} in which we seek to solve the dividend maximization problem and hence to compute the risk measure. That this method works also in high dimensions is demonstrated at the end of Section \ref{sec:dividends}, where numerical results are presented.

The method presented here can be applied to many other high-dimensional semilinear (degenerate) elliptic PDE problems in insurance mathematics, such as the calculation of ruin probabilities, but we emphasize that its application possibilities are not limited to insurance problems.

\section{BSDEs associated with elliptic PDEs}\label{sec:bsde-pde}

This section contains a short survey on scalar backward stochastic differential equations with random terminal times and on how they are related to a certain type of semilinear elliptic partial differential equations.

\subsection{BSDEs with random terminal times}
Let $(\Omega,\mathcal{F},\mathbb{P},(\cF_t)_{t\in[0,\infty)})$ be a filtered probability space satisfying the usual conditions and let
$W=(W_t)_{t\in[0,\infty)}$ be a $d$-dimensional standard Brownian motion on it. We assume that $(\cF_t)_{t\in[0,\infty)}$ is equal to the augmented natural filtration generated by $W$.
For all real valued row or column vectors $x$, let $|x|$ denote their Euclidean norm. We need the following notations and definitions for BSDEs.

\begin{defi}\label{BSDErtt}

A {\it BSDE with random terminal time} is a triple $(\tau,\xi,\f)$, where

\begin{itemize}
\item the {\it terminal time} $\tau\colon \Omega\to{[0,\infty]}$ is an $(\mathcal{F}_t)_{t\in[0,\infty)}$-stopping time, 

\item the {\it generator} $\f\colon\Omega\times{[0,\infty)}\times\mathbb{R}\times\mathbb{R}^{1\times d}\to\mathbb{R}$ is a process which satisfies that for all $(y,z)\in\mathbb{R}\times\mathbb{R}^{1\times d}$, the process $t\mapsto \f(t,y,z)$ is progressively measurable,

\item the {\it terminal condition} $\xi\colon\Omega\to\mathbb{R}$ is an $\mathcal{F}_\tau$-measurable random variable with $\xi=0$ on $\{\tau=\infty\}$.
\end{itemize}
\end{defi} 

\begin{defi}
 A {\it solution to the BSDE} $(\tau,\xi,\f)$ is a pair of progressively measurable processes $(Y,Z)=\big((Y_t)_{t\in[0,\infty)}, (Z_t)_{t\in[0,\infty)}\big)$ with values in $\R\times\R^{1\times d}$, where
\begin{itemize}
\item $Y$ is continuous $\mathbb{P}$-a.s.~and for all $T\in(0,\infty)$, the trajectories $t\mapsto{Z_t}$ belong to $L^2([0,T],\mathbb{R}^{1\times d})$, and $t\mapsto\f(t,Y_t,Z_t)$ is in $L^1([0,T])$,
\item for all $T\in(0,\infty)$ and all $t\in[0,T]$ it holds a.s.~that 
 \begin{align}\label{eq:bsde} 
Y_t=Y_T+\int_{t\wedge \tau}^{T\wedge \tau}\f(s,Y_s,Z_s)ds-\int_{t\wedge \tau}^{T\wedge \tau}Z_sd W_s,
\end{align}
\item $Y_t=\xi$ and $Z_t=0$ on $\{ t\geq \tau\}$.
\end{itemize}
\end{defi} 
Results on existence of solutions of BSDEs with random terminal time can be found in Pardoux' seminal article \cite{pardoux1998backward} (see {\cite[Theorem 3.2]{pardoux1998backward}}), in \cite{confortola2007backward} for generators with quadratic growth (see \cite[Theorem 3.3]{confortola2007backward}), and, e.g., in \cite{darling1997backward,briand1998backward,pardoux1999backward,briand2003backward,royer2004backward}; many of them cover multidimensional state spaces for the $Y$-process.\\

Optimal control problems which can be treated using a BSDE setting have for example been studied in \cite[Section 6]{confortola2007backward}. For this they consider generators of the forward-backward form
\begin{equation}\label{genF}
\f(t,y,z)=F(X_t,y,z)=\inf\{g(X_t,u)+zr(X_t,u)\colon u\in \mathcal{U}\}-\lambda y,
\end{equation}
where $X$ is a forward diffusion (see also the notation in the following subsection), $\mathcal{U}$ is a Banach space, $r$ is a Hilbert space-valued function (in their setting $z$ takes values in the according dual space) with linear growth, $g$ a real valued function with quadratic growth in $u$, and $\lambda\in(0,\infty)$.
In the sequel we focus on generators of forward-backward form.

\subsection{Semilinear elliptic PDEs and BSDEs with random terminal time}

In this subsection we recall the connection between semilinear elliptic PDEs and BSDEs with random (and possibly infinite) terminal time. The relationship between the theories is based on a nonlinear extension of the Feynman-Kac formula, see \cite[Section 4]{pardoux1998backward}.\\

We define the {\it forward process} $X$ as the stochastic process satisfying a.s.,
\begin{align}\label{eq:forward-SDE}
X_t=x+\int_0^t\mu(X_s)ds+\int_0^t\sigma(X_s)dW_s,\quad t\in[0,\infty),
\end{align}
where $x\in \mathbb{R}^d$ and $\mu\colon\mathbb{R}^d\to\mathbb{R}^d$ and $\sigma\colon\mathbb{R}^d\to\mathbb{R}^{d\times d}$ are globally Lipschitz functions.\\

In this paper we consider the following class of PDEs.
\begin{defi}
\begin{itemize}
\item A {\it semilinear (degenerate) elliptic PDE on the whole $\R^d$} is of the form
\begin{align}\label{eq:pdesyst}
\cL u+F(\cdot,u,(\nabla u)\sigma)=0,
\end{align}
where the differential operator $\cL$ acting on $C^2(\mathbb{R}^d)$ is given by
\begin{align}
\label{eq:Loperator}
\cL := \Loperator{},
\end{align}
and $F\colon\mathbb{R}^d\times\mathbb{R}\times\mathbb{R}^{1\times d}\to\mathbb{R}$ is such that the process $(t,y,z)\mapsto F(X_t,y,z)$ is a generator of a BSDE in the sense of Definition \ref{BSDErtt}.

\item We say that a function $u$ satisfies equation \eqref{eq:pdesyst} with {\it Dirichlet boundary conditions} on the open, bounded {\it domain} $G\subseteq\mathbb{R}^d$, if
\begin{equation}\label{eq:dirichlet}
\begin{aligned}
&\cL u+F(\cdot,u,(\nabla u)\sigma)=0,\quad x\in G,\\
&u(x)=g(x),\quad x\in\partial G,
\end{aligned}
\end{equation}
where $g\colon\mathbb{R}^d\to\mathbb{R}$ is a bounded, continuous function. 
\end{itemize}
\end{defi}

\begin{defi}
\begin{enumerate}
\item A {\it BSDE associated to the PDE \eqref{eq:pdesyst} on the whole $\mathbb{R}^d$} is given by the triplet $(\tau,\xi,\f)$, where $\tau\equiv \infty$, $\xi=0$, $\f(t,y,z)=F(X_t,y,z)$, $X$ is as in \eqref{eq:forward-SDE}, and
the solution satisfies a.s.~for all $T\in(0,\infty)$ that
\begin{align}\label{eq:bsde-pde}
Y_t=Y_T+\int_t^TF(X_s,Y_s,Z_s)ds-\int_t^TZ_sdW_s,\quad t\in[0,T].
\end{align}

\item A {\it BSDE associated to the PDE \eqref{eq:dirichlet} with Dirichlet boundary conditions} is given by the triplet $(\tau,g(X_{\tau}),\f)$, where $\tau=\inf\{t\in[0,\infty)\colon X_t\notin \overline{G}\}$, $\f(t,y,z)=F(X_t,y,z)$, $X$ is as in \eqref{eq:forward-SDE}, and
the solution satisfies a.s.~for all $T\in(0,\infty)$ that
\begin{equation}\label{eq:bsde-diri}
\begin{aligned}
&Y_t=Y_T+\int_{t\wedge \tau}^{T\wedge \tau}F(X_s,Y_s,Z_s)ds-\int_{t\wedge \tau}^{T\wedge \tau}Z_sdW_s,\quad t\in[0,T],\\
&Y_t=g(X_{\tau}), \,Z_t =0 , \quad t\ge\tau.
\end{aligned}
\end{equation}
\end{enumerate}
\end{defi}
In order to keep the notation simple, we do not highlight the dependence of $X, Y, Z$ on $x$.\\
 
For later use we also introduce the following notion of solutions of PDEs, which we will use later.
\begin{defi}
\begin{itemize}
\item A function $u\in C(\mathbb{R}^d)$ is called {\it viscosity subsolution} of \eqref{eq:pdesyst}, if for all $\varphi\in C^2(\mathbb{R}^d)$ and all points $x\in \mathbb{R}^d$ where $u-\varphi$ has a local maximum,
$$\cL\varphi(x)+F(x,u(x),(\nabla \varphi(x))\sigma(x))\ge 0.$$
\item A function $u\in C(\mathbb{R}^d)$ is called {\it viscosity supersolution} of \eqref{eq:pdesyst}, if for all $\varphi\in C^2(\mathbb{R}^d)$ and all points $x\in \mathbb{R}^d$ where $u-\varphi$ has a local minimum,
$$\cL\varphi(x)+F(x,u(x),(\nabla \varphi(x))\sigma(x))\le0.$$
\item A function $u\in C(\mathbb{R}^d)$ is called {\it viscosity solution} of \eqref{eq:pdesyst}, if it is a viscosity sub- and supersolution.
\end{itemize}
\end{defi}
A similar definition of viscosity solutions can be given for the case of Dirichlet boundary conditions \eqref{eq:dirichlet}, see \cite{pardoux1998backward}.\\

For later use, note that \eqref{eq:bsde-diri} can be rewritten in forward form as
\begin{equation}\label{forw-Y}
\begin{aligned}
&Y_t=Y_0-\int_0^t F(X_s,Y_s,Z_s)ds+\int_0^tZ_sdW_s,\quad t\in[0,\tau),\\
&Y_t=g(X_{\tau}), \,Z_t =0 , \quad t\ge\tau.
\end{aligned}
\end{equation}

The following theorems link the semilinear elliptic PDEs \eqref{eq:pdesyst} and \eqref{eq:dirichlet} to the associated BSDEs.

\begin{theorem}[{\cite[Theorem 4.1]{pardoux1998backward}}]
	
	Let $(t,y,z)\mapsto F(X_t,y,z)$ meet the assumptions of 
	 \cite[Theorem 3.2]{pardoux1998backward} 
	and let $u \in C^2(\Rd)$ satisfy
	$$
	\E\!\left[ \int_0^\infty e^{\lambda t} | ((\grad u) \sigma)(X_t) |^2 dt\right] < \infty
	$$
	with $\lambda$ as in \cite[Theorem 3.2]{pardoux1998backward}.
	 If $u$ is a classical solution of the PDE \eqref{eq:pdesyst},	
	then
	$$
	Y_t = u(X_t), \quad Z_t = ((\grad u) \sigma)(X_t)
	$$
	solve the BSDE \eqref{eq:bsde-pde}. An equivalent statement can be established for the system with boundary conditions \eqref{eq:dirichlet} and equation \eqref{eq:bsde-diri}, see \cite{pardoux1998backward}. 
\end{theorem}
Note that for all $x\in \R^d$, $Y$ and $Z$ are stochastic processes adapted to $(\cF_t)_{t\in[0,\infty)}$. Therefore $Y_0$, $Z_0$ are $\cF_0$-measurable and hence a.s.~deterministic.
For us, the connection between PDEs and BSDEs is of relevance because of the converse result,
where $x\mapsto Y_0$ delivers a solution to the respective PDE.
\begin{theorem}[{\cite[Theorem 4.3]{pardoux1998backward}}]\label{thm:bsdepde1}
Assume that for some $K,K',p\in(0,\infty)$, $\gamma\in(-\infty,0)$ the function $F$ satisfies for all $x,y,y',z,z',$
\begin{enumerate}[label=(\roman*)]
\item $|F(x,y,z)|\leq K'(1+|x|^p+|y|+|z|)$,
\item $\langle y-y',F(x,y,z)-F(x,y',z)\rangle\leq \gamma|y-y'|^2$,
\item $|F(x,y,z)-F(x,y,z')|\leq K|z-z'|$.
\end{enumerate}
Then \cite[Theorem 3.2]{pardoux1998backward} can be applied to the generator $(t,y,z)\mapsto F(X_t,y,z)$, showing that the function $u$ given by $u(x)=Y_0$ is a viscosity solution to \eqref{eq:pdesyst}, where $Y$ is the first component of the unique solution to \eqref{eq:bsde-pde} in the class of solutions from \cite[Theorem 3.2]{pardoux1998backward}.
\end{theorem}
The case of the Dirichlet problem requires additional assumptions on the domain $G$ and the exit time $\tau$ from \eqref{eq:bsde-diri}. We refer to \cite[Theorem 4.3]{pardoux1998backward}. A corresponding result for BSDEs with quadratic generator is \cite[Theorem 5.2]{confortola2007backward}.\\

To conclude, the correspondence between PDE \eqref{eq:dirichlet} and BSDE \eqref{eq:bsde-diri} is given by $Y_t=u(X_t)$, $Z_t=((\nabla u)\sigma)(X_t)$, $\xi=g(X_\tau)$. 
For tackling elliptic PDEs which are degenerate (as it is the case for our insurance mathematics example) we need to take the relationship a little further in order to escape the not so convenient structure of the $Z$-process.
We factor $\mathcal Z \sigma(X) = Z$ for cases where this equation is solvable for $\mathcal Z$ ($\sigma$ needs not necessarily be invertible) and define $f(x,y,\zeta):=F(x,y,\zeta \sigma(x))$\footnote{Since $\mathcal Z \sigma(X) = Z$ is solvable for $\mathcal Z$, $f$ is well-defined.}, giving the correspondence $Y_t=u(X_t)$, $\mathcal{Z}_t=\nabla u(X_t)$, $\xi=g(X_\tau)$.
This relationship motivates us to solve semilinear degenerate elliptic PDEs by solving the corresponding BSDEs forward in time (cf.~\eqref{forw-Y})
\begin{align*}
Y_t=Y_0-\int_0^t f(X_s,Y_s,\mathcal Z_s)ds+\int_0^t\mathcal Z_s \sigma(X_s) dW_s,  \quad t\in[0,\tau)
\end{align*}
for $Y_0$ by approximating the paths of $\mathcal Z=\nabla u(X)$ by a DNN, see Section \ref{sec:algorithm}. Doing so, we obtain an estimate of a solution value $u(x)$ for a given $x \in \Rd$.

\section{Algorithm}\label{sec:algorithm}

The idea of the proposed algorithm is inspired by \cite{han2018solving}, where the authors use the correspondence between BSDEs and semilinear parabolic PDEs to construct a DNN algorithm for solving the latter. In the same spirit, we construct a DNN algorithm based on the correspondence to BSDEs with random terminal time for solving semilinear elliptic PDEs. 

The details of the algorithm are described in three steps of increasing specificity. First we explain the DNN algorithm mathematically. This is done below. Second, Algorithm 1 at the end of this section provides a pseudocode. Third, our program code is provided on Github\footnote{https://github.com/stefankremsner/elliptic-pdes} under a creative commons license.
The algorithm is implemented in a generic manner so that it can be reused for other elliptic PDE problems.\\

The goal of the algorithm is to calculate solution values $u(x)$ of the semilinear (degenerate) elliptic PDE of interest.
For the construction of the algorithm we use the correspondence to a BSDE with random terminal time. Recall from Section \ref{sec:bsde-pde} that such a BSDE is given by a triplet $(\tau, g(X_\tau), f)$ that can be determined from the given PDE and by
\begin{align}\label{forw1}
X_t=x+\int_0^t\mu(X_s)ds+\int_0^t\sigma(X_s)dW_s
\end{align}
and
\begin{align}\label{backw1}
Y_t=Y_0-\int_0^t f(X_s,Y_s,\mathcal Z_s)ds+\int_0^t\mathcal Z_s \sigma(X_s) dW_s.
\end{align}
Furthermore, recall that we have identified $Y_t=u(X_t)$, where $u$ is the solution of the PDE we are interested in.\\

The first step for calculating $u$ is to approximate \eqref{forw1} up to the stopping time $\tau$.
To make this computationally feasible, we choose $T$ large and stop at $\tau\wedge T$, hence at $T$ at the latest.
Now let $0= t_0 < t_1 < \dots < t_N=T$, $\Delta t_n = t_{n+1} - t_n$.
We simulate $M$ paths $\omega_1,\dots,\omega_M$ of the Brownian motion $W$. With this we
approximate the forward process using the Euler-Maruyama scheme, that is $X_0=x$ and
\begin{align}
X_{t_{n+1}} \approx X_{t_n} + \mu(X_{t_n}) \Delta t_n + \sigma(X_{t_n}) \Delta W_n.
\end{align}

In the next step we compute $\mathcal Z$. For all $t_n$, $\mathcal Z_{t_n}=\grad u(\Xtn)$ are approximated by DNNs, each mapping $G$ to $\R^d$.
As noted above, the implementation of this (and all other steps) is provided.\\

Now, we initialize $u(x)$ and use the above approximations to compute the solution to the BSDE forward in time by approximating \eqref{backw1}:
\begin{equation}
\label{eq:euler-maruyama-scheme}
\begin{aligned}
u(\Xtnplus) &\approx  u(\Xtn) - \ind_{(0,\tau)}(t_n) f\left(\Xtn, u(\Xtn), \grad u(\Xtn) \right) \Delta t_n \\
&\quad +  \ind_{(0,\tau)}(t_n) \grad u(\Xtn) \sigma(\Xtn) \Delta W_n.
\end{aligned}
\end{equation}

Note that due to this construction, indirectly $u(\Xtnplus)$ is also approximated by a DNN as a combination of DNNs.

For the training of the involved DNNs, we compare $u(X_{\tau\wedge T})$ with the terminal value $\xi$. This defines the loss function for the training:
$$\frac{1}{M} \sum_{k=1}^M |u(X_{\tau \wedge T}(\omega_k)) - \xi(\omega_k)|^2.$$
After a certain number of training epochs the loss function is minimized and we obtain an approximate solution value $u(x)$ of the PDE.\\

\begin{rem}
Several approximation errors arise in the proposed algorithm:
\begin{enumerate}
	\item the approximation error of the Euler-Maruyama method, which is used for sampling the forward equation,
	\item the error of approximating the expected loss,
	\item the error of cutting off the potentially unbounded random terminal time at time $T$,
	\item the approximation error of the deep neural network model for approximating $\mathcal Z_{t_n}$ for each $t_n$.
	\end{enumerate}
It is well known that for any continuous function we can find DNNs that approximate the function arbitrarily well, see \cite{hornik1,hornik2}. This is, however, not sufficient to make any statement about the approximation quality. Results on convergence rates are required.
 Though this question is already studied in the literature (see, e.g., 
	\cite{beck2020overcomingelliptic,	
		AllenCahnPaper,		
		BernerGrohsJentzen2018,
		ElbraechterSchwab2018,
		GononGrohsJentzenKoflerSiska2019Uniform,
		GrohsWurstemberger2018,
		GrohsHornungJentzen2019,
		grohs2019deep,
		HutzenthalerJentzenKruse2019,
		Overcoming,
		hutzenthaler2019overcoming,
		JentzenSalimovaWelti2018,
		KutyniokPetersen2019,
		ReisingerZhang2019}), 
	results on convergence rates for given constructions are yet scarce and hence many questions remain open while the number of proposed DNN algorithms grows.
\end{rem}

We close this section with some comments on the implementation.
\begin{rem}
\begin{itemize}
\item All DNNs are initialized with random numbers. 
\item For each value of $x$ we average $u(x)$ over 5 independent runs. The estimator for $u(x)$ is calculated as the mean value of $u(x)$ in the last 3 network training epochs of each run, sampled according to the validation size (see below).
\item We choose a non-equidistant time grid in order to get a higher resolution for earlier (and hence probably closer to the stopping time) time points.
\item We use  $\tanh$ as activation function.
\item We compute $u(x)$ simultaneously for 8 values of $x$ by using parallel computing.
\end{itemize}
\end{rem}

\begin{algorithm}
	\caption{Elliptic PDE Solver for a BSDE $(f, \xi)$ with stopping time $\tau$}
	{\fontsize{10}{10}\selectfont
		\begin{algorithmic}[1]
			\Require{number of training epochs $E$, maximal time $T$, step-size $\Delta t$,  number of timesteps $N$, number of sample paths $M$, number of hidden layer neurons dim, initial (random) starting values $(\theta_0^{(u)}, \theta_0^{(\zeta)})$ }\vspace{3mm}
			
			\Function{TrainableVariables}{$\text{dim},\theta$}
			\Comment{see Pytorch or Tensorflow}
			
			\Return a trainable variable with dimension $1 \times \text{dim}$ initialized by $\theta$.
			
			\EndFunction\vspace{3mm}
			
			\Function{Subnetwork}{$x$}
			\Comment{allowing $x$ to be a tensor containing $M$ rows (samples)}
			
			\Return a trainable DNN, evaluated at $x$.				
			
			\EndFunction
			
			\Statex
			
			\For{$i = 0,\dots,N$} 
			\State $t_i = \text{timesteps}(i)$
			\Comment{Initialize non-equidistant timesteps}
			\EndFor
			
			\Statex
			
			\For{$j = 1,\dots,M$} 
			
			\State Sample Brownian motion trajectory $\left(w_{t_i}^{(j)}\right)_{0 \le i \le N}$ 
			
			\State Sample path from forward process $\left(x_{t_i}^{(j)}\right)_{0 \le i \le N}$
			
			\State calculate stopping time $\tau^{(j)}$
			\State calculate terminal value $\xi^{(j)}$ 
			\State set $x_t^{(j)} = x^{(j)}_{\tau^{(j)}}$ for all $t > \tau^{(j)}$ 
			\EndFor\vspace{3mm}
			
			\State $u_0 = \Call{TrainableVariables}{1,\theta_0^{(u)}$}  \Comment{Initialize $u$}
			
			\State $\grad u_0 = \Call{TrainableVariables}{d,\theta_0^{(\zeta)}$}   \Comment{Initialize $\mathcal Z$}
			
			\For{$j = 1,\dots,M$} 		
			\State $u^{(j)} = u_0$ 
			\State $\grad u^{(j)} = \grad u_0$	
			\EndFor
			
			\For{$e = 1,\dots,E$} 	
			\For{$i = 1,\dots,N-1$} 				
			\For{$j = 1,\dots,M$} 	
			\State  $u^{(j)} = u^{(j)} - f(x_{t_i}^{(j)},u^{(j)}, \grad u^{(j)}) (t_{i+1}-t_i) + \grad u^{(j)} \sigma(x^{(j)}_{t_i})(w_{t_{i+1}}^{(j)}-w_{t_i}^{(j)})$
			\If{$t_{i+1} > \tau^{(j)}$} 
			break
			\EndIf			
			\EndFor
			\State $\grad u = \Call{Subnetwork}{x_{t_{i+1}}}$
			\EndFor 
			\State update all trainable variables and the subnetwork's weights according to the loss function
			$$\frac{1}{M}\sum_{j=1}^M(u^{(j)} - \xi^{(j)})^2$$
			\EndFor
			
			\Return $(u_0, \grad u_0)$
			
		\end{algorithmic}
	}
\end{algorithm}

\section{Examples}

In this section we apply the proposed algorithm to three examples. The first one serves as a validity check, the second one as an academic example with a non-linearity. Finally, we apply the algorithm to solve the dividend maximization problem under incomplete information.

\subsection{The Poisson equation}
\label{sec:Poisson}

The first example we study is the Poisson equation -- a linear PDE.

Let $r\in(0,\infty)$, $G=\left\{x \in \Rd \colon |x| < r \right\}$, $\partial G=\left\{x \in \Rd \colon |x| = r \right\}$, $b\in\mathbb{R}$, and
\begin{equation}\label{eq:simple_poisson_equation}
\begin{aligned}
		\laplace u(x) = -b, & \quad x \in G, \\
	u(x) = 0, &\quad x \in \partial G .
 \end{aligned}
  \end{equation}
Solving \eqref{eq:simple_poisson_equation} is equivalent to solving the BSDE with
\begin{equation*}
\begin{aligned}
  dX_t &= \sqrt{2}dW_t, \qquad X_0 = x, \\
f(x,y,\zeta) &= b, \qquad \xi = 0 ,
\end{aligned}
  \end{equation*}
up to the stopping time $\tau=\inf\{t\in[0,T]\colon  |x| > r\}$.

To obtain a reference solution for this linear BSDE we use an analytic formula for the expectation of $\tau$, see \cite[Example 7.4.2, p.~121]{oksendal2003stochastic}. This yields
$$
u(x) 
= \frac{b}{2d}\left(r^2 - |x|^2\right) .
$$

\subsubsection{Numerical results}

\begin{figure}
	\centering
	\includegraphics[width=0.45\textwidth]{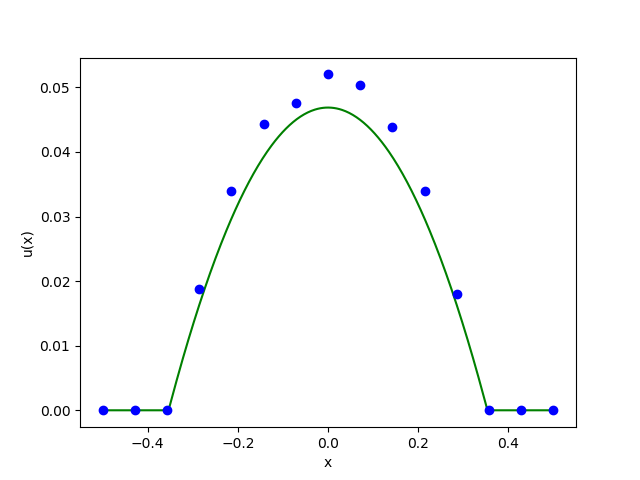}
	\includegraphics[width=0.45\textwidth]{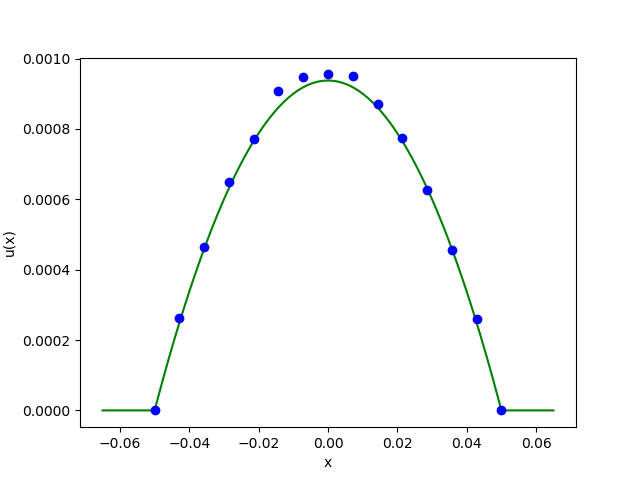}
			\caption{Approximate solution (blue) and reference solution (green) for the Poisson equation on the $\R^2$ (left) and on the $\R^{100}$ (right).}
	\label{fig:laplace}
\end{figure}

We compute $u(x)$ on the $\R^2$ and the $\R^{100}$ for 15 different values of $x$. Figure \ref{fig:laplace} shows the approximate solution of $u$ obtained by the DNN algorithm on the diagonal points $\{(x,\dots,x)\in\R^d \colon x \in [-r, r]\}$ (in blue) and the analytical reference solution (in green).
Table \ref{table1} contains the parameters we use.

\begin{table}
	\begin{center}
		\begin{tabular}{|c|c|c|c|c|c|c|c|c|}
			\hline
			$d$ &  $r$ &  $b$ & $N$ & $T$ & E & M & validation size & time per eight points\footnotemark\\
			\hline
			
			2 & 0.5 & $0.75$ & 500 & 0.5       & 200 & 64 & 256 & 119.17 s \\
			\hline
			100 & 0.5 & $0.75$ & 500 & 0.01 & 200 & 64 & 256 & 613.86 s  \\
			\hline
		\end{tabular}
	\end{center}
	\caption{\label{table1}Parameters for the Poisson equation.}
\end{table}

Note that as the expected value of $\tau$ decreases linearly in $d$, we adapt the cut off time $T$ for $d=100$ accordingly.

\footnotetext{The numerical examples were run on a Lenovo Thinkpad notebook with an Intel Core i7 processor (2.6 GHz) and 16 GB memory, without CUDA.}

\subsection{Quadratic gradient}
\label{sec:quadratic}

The second example is a semilinear PDE with a quadratic gradient term.

Let $r\in(0,\infty)$, $G=\left\{x \in \Rd \colon |x| < r \right\}$, and $\partial G=\left\{x \in \Rd \colon |x| = r \right\}$. We consider the PDE
\begin{equation}\label{eq:quadraticz_pde}
\begin{aligned}
\laplace u(x) + |\nabla u(x)|^2 = 2 e^{-u(x)}, & \quad x \in G, \\
u(x) = \log\!\left(\frac{r^2+1}{d}\right), &\quad x \in \partial G .
\end{aligned}
\end{equation}
corresponding to the BSDE
\begin{align}
\label{eq:pde_quadratic_gradient}
dX_t &= \sqrt{2}dW_t, \qquad X_0 = x, \\
f(x,y,\zeta) &=| \zeta |^2 - 2 e^{-y}, \qquad
\xi = \log\left(\frac{|r|^2+1}{d}\right).
\end{align}
In addition, this example we have an analytic reference solution given by
\begin{align*}
u(x) &= \log \left( \frac{|x|^2 + 1}{d} \right).
\end{align*}

\subsubsection{Numerical results}

\begin{figure}
	\centering
	\includegraphics[width=0.45\textwidth]{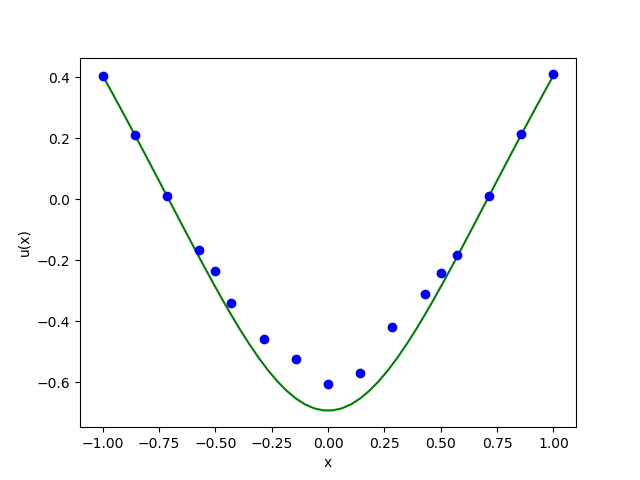}
		\includegraphics[width=0.45\textwidth]{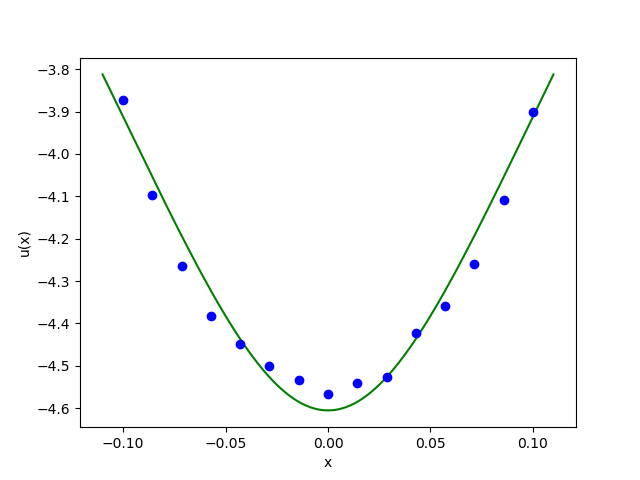}
		\caption{Approximate solution (blue) and reference solution (green) for the equation with quadratic gradient on the $\R^2$ (left) and on the $\R^{100}$ (right).}
	\label{fig:quadraticz}
\end{figure}

As in the previous example we compute $u(x)$ for 15 different values of $x$ on the $\R^2$ and the $\R^{100}$. Figure \ref{fig:quadraticz} shows the approximate solution of $u$ obtained by the DNN algorithm on the diagonal points $\{(x,\dots,x)\in\R^d \colon x \in [-r, r]\}$ (in blue) and the analytical reference solution (in green).
Table \ref{table2} contains the parameters we use.

\begin{table}
	\begin{center}
		\begin{tabular}{|c|c|c|c|c|c|c|c|c|}
			\hline
			$d$ &  $r$ & $N$ & $T$ & E & M & validation size & time per eight points\\
			\hline
			2 &  1 & 100 & 5   & 500 & 64 & 256 & 204.58 s  \\
			\hline
			100 &  1 & 100 & 0.1 & 500 & 64 & 256 &  321.13 s  \\
			\hline
		\end{tabular}
	\end{center}
	\caption{\label{table2}Parameters for the equation with quadratic gradient.}
\end{table}

While classical numerical methods for PDEs would be a much better choice in the case $d=2$, their application would not be feasible in the case $d=100$.

\subsection{Dividend maximization}
\label{sec:dividends}

The goal of this paper was to show how to use the proposed DNN algorithm to solve high-dimensional control problems that arise in insurance mathematics. We finally arrived at the point where we are ready to do so.\\

Our example comes from \cite{S16}, where the author studies De Finetti's dividend maximization problem in a setup with incomplete information about the current state of the market. The hidden market-state process determines the trend of the surplus process of the insurance company and is modeled as a $d$-state Markov chain. Using stochastic filtering, in \cite{S16} they achieve to transform the one-dimensional problem under incomplete information to a $d$-dimensional problem under complete information. The cost is $(d-1)$ additional dimensions in the state space.
We state the problem under complete information using different notation than in \cite{S16} in order to avoid ambiguities.

The probability that the Markov chain modeling the market-state is in state $i\in\{1, \dots, d-1\}$ is given by
\begin{align}
\label{eq:dynpi}\pi_i(t) = x_i +\int_0^t \left(q_{d,i} + \sum_{j=1}^{d-1} (q_{j,i}-q_{d,i}) \pi_j (s)\right) \, ds 
+ \int_0^t \pi_i (s) \frac{\trend_i-\nu_s}{\rho} \, dB_s,
\end{align}
where
\begin{align}\label{eq:nu}
\nu_t=\trend_d + \sum_{j=1}^{d-1}(\trend_j-\trend_d)\pi_j(t),
\end{align} 
$x_i \in(0,1)$, $B$ is a one-dimensional Brownian motion, $\trend_1,\dots,\trend_d\in \R$ are the values of the surplus trend in the respective market-states of the hidden Markov chain, and $(q_{i,j})_{i,j\in\{1,\dots,d\}}\in\R^{d\times d}$ denotes the intensity matrix of the chain.

Let $(\ell_t)_{t\in[0,\infty)}$ be the dividend rate process. The surplus of the insurance company is given by
\begin{align}
\label{eq:dynX2}\widetilde X^d_t = x_d + \int_0^t (\nu_s-\ell_s) \,ds + \rho B_t, \quad t\in[0,\infty),
\end{align}
where $x_d,\rho\in(0,\infty)$.
For later use we define also the dividend free surplus
\begin{align}
\label{eq:Z0}X^d_t = x_d + \int_0^t \nu_s \,ds + \rho B_t, \quad t\in[0,\infty).
\end{align}
The processes \eqref{eq:dynpi} and \eqref{eq:dynX2} form the $d$-dimensional state process underlying the optimal control problem we aim to solve.

The goal of the insurance company is to determine its value by maximizing the discounted dividends payments until the time of ruin $\eta=\inf\{t\in(0,\infty]\colon \widetilde X^d_t<0\}$, that is it seeks to find
\begin{align}\label{divprob}
u(x_1,\dots,x_{d})=\sup_{(\ell_t)_{t\in[0,\infty)}\in A}  \E_{x_1,\dots,x_{d}} \!\left[\int_0^{\eta}e^{-\delta t}\ell_t \,dt\right],
\end{align}
where $\delta\in(0,\infty)$ is a discount rate, $A$ is the set of admissible controls, and $\E_{x_1,\dots,x_{d}}[\cdot]$ denotes the expectation under the initial conditions $\pi_i(0)=x_i$ for $i\in\{1,\dots,d-1\}$ and $\widetilde X^d_0=x_d$. Admissible controls are $(\cF^{X^d}_t)_{t\ge 0}$-progressively measurable, $[0,K]$-valued for $K\in(0,\infty)$, and fulfill $\ell_t \equiv 0$ for $t>\eta$, cf.~\cite{S16}.\\

In order to tackle the problem, we solve the corresponding Hamilton-Jacobi-Bellmann (HJB) equation\footnote{For abbreviation we use $u_{x_d}$ for $\frac{\partial u}{\partial x_d}$ etc.} from \cite{S16},
\begin{align}
\label{eq:HJB}(\cL - \delta) u +\sup_{\ell \in [0,K]}(\ell(1-u_{x_d}))=0,
\end{align}
where $\cL$ is the second order degenerate elliptic operator
\begin{align*}
\cL u&=  \trend_d u_{x_d}+\sum_{i=1}^{d-1} \left((\trend_i-\trend_d) x_i u_{x_d} + \left(q_{di} + \sum_{j=1}^{d-1} (q_{ji}-q_{di}) x_i  \right) u_{x_i} + x_i \left( \trend_i - \nu \right) u_{x_d x_i}\right.\\
&+ \left.\frac{1}{2}\sum_{j=1}^{d-1} \left( \left( x_i \frac{\trend_i-\nu}{\rho} \right) \left( x_j \frac{\trend_j-\nu}{\rho} \right) u_{x_i x_j} \right)\right)+ \frac{1}{2} \rho^2 u_{x_d x_d}.
\end{align*}
The supremum in \eqref{eq:HJB} is attained at
\begin{align*}
\ell=\begin{cases}
K, & u_{x_d} \le 1\\
0, & u_{x_d} > 1.
\end{cases}
\end{align*}
Plugging this into \eqref{eq:HJB} we end up with a $d$-dimensional semilinear degenerate elliptic PDE:
\begin{align}
\label{eq:HJB_neu}(\cL - \delta) u +K(1-u_{x_d})\ind_{\{u_{x_d}\le 1\}} =0.
\end{align}
The boundary conditions in $x_d$ direction are given by
\begin{align*}
u(x_1,\dots,x_d)=\begin{cases}
K/\delta, & x_d \to\infty\\
0, & x_d=0.
\end{cases}
\end{align*}
No boundary conditions are required for the other variables, cf.~\cite{S16}.\\

In \cite[Corollary 3.6]{S16} it is proven that the unique viscosity solution to \eqref{eq:HJB_neu} solves the optimal control problem \eqref{divprob}. Hence, we can indeed solve the control problem by solving the HJB equation.\\

For the numerical approximation we cut off $x_d$ at $r\in(0,\infty)$. Hence,
$G=\left\{x \in \Rd \colon 0 < x_d <r\right\}$ and $\partial G=\left\{x \in \Rd \colon x_d \in\{0,r\} \right\}$.

For the convenience of the reader we derive the BSDE corresponding to \eqref{eq:HJB_neu}.
The forward equation is given by
$$
dX_t = (d\pi_1(t),\dots,d\pi_{d-1}(t),dX^d_t)^\top, \qquad X_0 = x, \\
$$
that is 
$$
dX_t = \mu(X_t) dt + \sigma(X_t) dW_t, \qquad X_0 = x, \\
$$
where $W=(B,W^2,\dots,W^d)^\top$, $x=(x_1,\dots,x_d)^\top$, and
\begin{align*}
\mu(x) = 
\left( 
q_{d,1} + \sum_{j=1}^{d-1} (q_{j,1}-q_{d,1})x_j, \dots, q_{d,d-1} + \sum_{j=1}^{d-1} (q_{j,d-1}-q_{d,d-1})x_{d-1},\trend_d + \sum_{j=1}^{d-1}(\trend_j-\trend_d)x_j
\right) ^\top,
\end{align*}
\begin{align*}
\sigma(x)= 
\left( 
\begin{array}{cccc}
  x_1 \frac{\trend_1-\trend_d + \sum_{j=1}^{d-1}(\trend_j-\trend_d)x_j}{\rho} & 0 & \dots & 0 \\
  \dots & 0 & \dots & 0\\
    x_{d-1} \frac{\trend_{d-1}-\trend_d + \sum_{j=1}^{d-1}(\trend_j-\trend_d)x_j}{\rho}  &  0 & \dots & 0\\
\rho &  0 & \dots & 0\\
\end{array}
\right) .
\end{align*}
We claim that the BSDE associated to \eqref{eq:HJB_neu} is given  in forward form by
\begin{equation}\label{BSDE-div}
\begin{aligned}
u(X_t)&=u(x)-\int_0^t [K(1-u_{x_d}(X_s))\ind_{\{u_{x_d}(X_s)\le 1\}}-\delta u(X_s) ] dt
\\&\quad + \int_0^t  \nabla u(X_s) \sigma(X_s)dW_s.
\end{aligned}
\end{equation}
Applying It\^o's formula to $u(X)$ yields
\begin{equation}\label{BSDE-div-ito}
\begin{aligned}
u(X_t)&=u(x)+\int_0^t \cL u(X_s) dt
+ \int_0^t \nabla u(X_s) \sigma(X_s) dW_s.
\end{aligned}
\end{equation}
Combining \eqref{BSDE-div} and \eqref{BSDE-div-ito} gives
\begin{align*}
&u(x)-\int_0^t [K(1-u_{x_d}(X_s)) \ind_{\{u_{x_d}(X_s)\le 1\}}-\delta u(X_s)] dt
 + \int_0^t \nabla u(X_s) \sigma(X_s) dW_s
\\&\quad=
u(x)+\int_0^t \cL u(X_s) dt
+ \int_0^t \nabla u(X_s) \sigma(X_s)  dW_s.
\end{align*}
Canceling terms verifies (in a heuristic manner) \eqref{eq:HJB_neu}. 

Hence, the corresponding BSDE has the parameters
\begin{align}
\label{eq:pde_div}
f(x,y,\zeta) &= 
 K(1-\zeta_d) \ind_{\{\zeta_d\le 1\}}-\delta y 
\end{align}
and
\begin{align*}
\xi=\begin{cases}
K/\delta, & X^d_\tau = r,\\
0, & X^d_\tau=0,
\end{cases}
\end{align*}
if $\tau<\infty$.

\subsubsection{Numerical results}

As for this example we have no analytic reference solution at hand, we use the solution from \cite{S16} for the case $d=2$, which was obtained by a finite difference method and policy iteration. Then we show that the DNN algorithm also provides an approximation in high dimensions in reasonable computation time.

\begin{figure}
	\centering
	\includegraphics[width=0.45\textwidth]{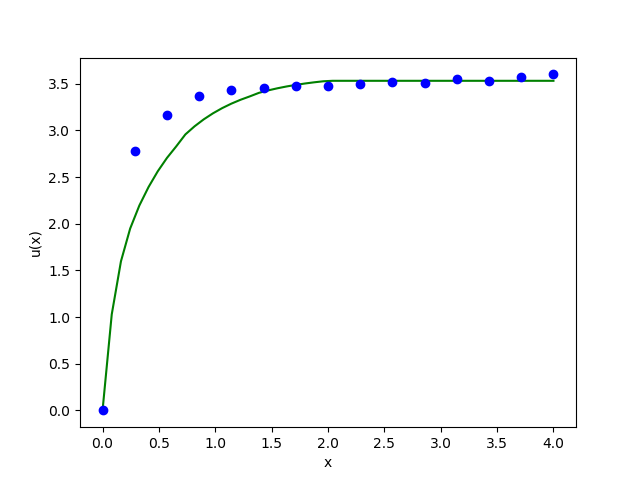}
	\includegraphics[width=0.45\textwidth]{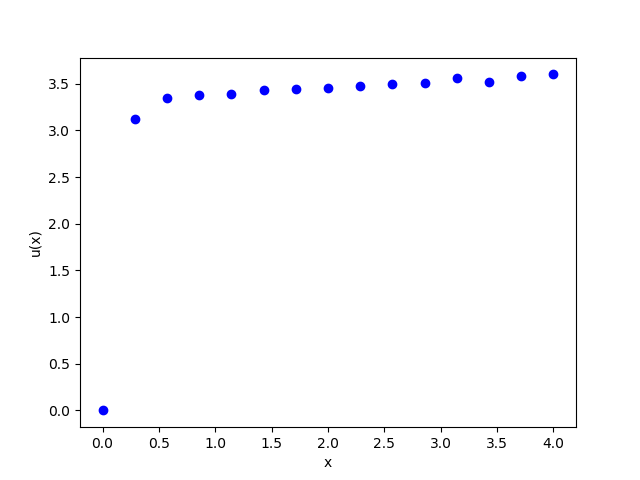}
		\caption{Approximate solution (blue) and reference solution (green) for the dividend problem on the $\R^2$ for fixed $\pi_1=\pi_2=0.5$ (left) and on the $\R^{100}$ for fixed $\pi_1=\dots=\pi_{100}=0.01$ (right, without reference solution).}
	\label{fig:insurrance-bsde}
\end{figure}

\begin{figure}
	\centering
	\includegraphics[width=\imagewidth]{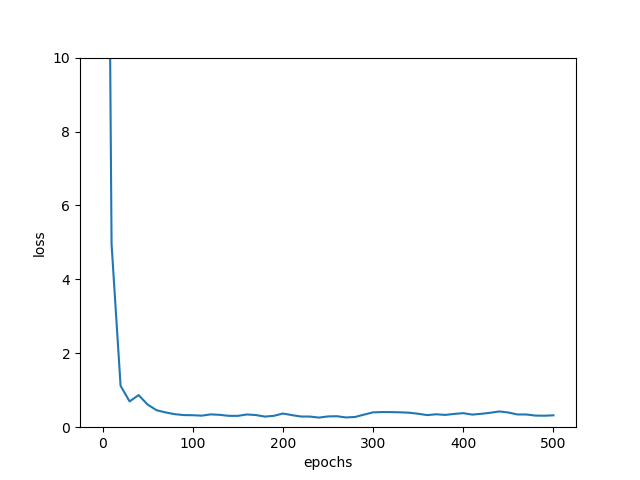}
		\caption{Interpolated loss for the case $d=100$.}
	\label{fig:loss}
\end{figure}

As in the previous examples we compute $u(x)$ on the $\R^2$ and on the $\R^{100}$ for 15 different values of $x$. Figure \ref{fig:insurrance-bsde} shows the approximate solution of the HJB equation and hence the value of the insurance company obtained by the DNN algorithm (in blue) and the reference solution from \cite{S16} (in green) for the case $d=2$.
For $d=100$ we have no reference solution at hand.  
Figure \ref{fig:loss} shows the loss for a fixed value of $x$ in the case $d=100$.
Tables \ref{table3} and \ref{table4} contain the parameters we use.

\begin{center}
	\begin{table}
		\begin{tabular}{|c|c|c|c|c|c|c|c|c|c|c|c|}
			\hline
			$d$ & $r$ & $K$ & $\delta$ & $\rho $ & $a_i$ & $N$ & $T$ & E & M & validation size & time per eight points\\
			\hline
			
			2 & 5 & 1.8 & 0.5 & 1 & $\left(2-\frac{i}{d}\right)$ & 100 & 5 & 500 & 64 & 256 & 317.42 s  \\
			\hline
			100 & 5 & 1.8 & 0.5 & 1 & $\left(2-\frac{i}{d}\right)$ & 100 & 5 & 500 & 64 & 256 & 613.15 s  \\
			\hline
		\end{tabular}
		\caption{\label{table3}{Parameters for the dividend problem.}}
	\end{table}
\end{center}

\begin{center}
	\begin{table}
		\begin{tabular}{|c|c|c|c|c|c|c|}
			\hline
			case & $i=j$ even & $i=j$ odd & $i=j+1$ even & $i=j+1\ge3$ odd & $i=1$, $j=d$ & otherwise \\
			\hline
			$q_{i,j}$& $-0.5$ & $-0.25$ & $0.5$ & 0.25 & 0.25 & 0\\
			\hline
		\end{tabular}
		\caption{\label{table4}{Intensity matrix values for the dividend problem.}}
	\end{table}
\end{center}

\section{Conclusion}

The goal of this paper was to compute the risk measure given by the expected discounted future dividend payments in a complex high-dimensional economic environment. This demonstrates the effectiveness of using DNN algorithms for solving some high-dimensional PDE problems in insurance mathematics that cannot be solved by classical methods. In the literature the focus so far was on parabolic PDE problems; however, in insurance mathematics we often face problems up to an unbounded random terminal time, e.g., the time of ruin of the insurance company, leading to (degenerate) elliptic problems.

Hence, we have proposed a novel deep neural network algorithm for a large class of semilinear (degenerate) elliptic PDEs associated to infinite time horizon control problems in high dimensions.
The method extends the DNN algorithm proposed by Han, Jetzen, and E \cite{han2018solving}, which was developed for parabolic PDEs, to the case of (degenerate) elliptic semilinear PDEs. We have attacked the problem inspired by a series of results by Pardoux \cite{pardoux1998backward}. 

Of course, in low dimensions one would not use the proposed DNN algorithm -- classical methods are more efficient.
However, recent models are frequently high dimensional, in which case classical methods fail due to the curse of dimensionality. Then the DNN algorithm presented here can be applied to compute the desired quantity.

We emphasize that the method presented here can also be applied to many other high-dimensional semilinear (degenerate) elliptic PDE problems in insurance mathematics and beyond.

An implementation of the algorithm is provided on Github\footnote{https://github.com/stefankremsner/elliptic-pdes} under a creative commons license.

\section*{Acknowledgements}
The authors thank Gunther Leobacher for discussions and suggestions that improved the paper.\\

S.~Kremsner is supported by the Austrian Science Fund (FWF): Project F5508-N26, which is part of the Special Research Program ``Quasi-Monte Carlo Methods: Theory and Applications''.

\bibliography{bib/full}{}
\bibliographystyle{plain}

\end{document}